\documentclass{article}
\usepackage{ijcai18}

\usepackage{times}
\usepackage{soul}
\usepackage{url}
\usepackage[hidelinks]{hyperref}
\usepackage[utf8]{inputenc}
\usepackage[small]{caption}
\usepackage{graphicx}

\title{Explainable Security}

\author{
Luca Vigan\`o, \
Daniele Magazzeni
\\ 
Department of Informatics \\
King's College London, UK \\
luca.vigano@kcl.ac.uk, daniele.magazzeni@kcl.ac.uk
}

\begin{document}

\maketitle

\begin{abstract}
The Defense Advanced Research Projects Agency (DARPA) recently launched the Explainable Artificial Intelligence (XAI) program that aims to create a suite of new AI techniques that enable end users to understand, appropriately trust, and effectively manage the emerging generation of AI systems.

In this paper, inspired by DARPA's XAI program, we propose a new paradigm in security research: Explainable Security (XSec).
We discuss the ``Six Ws'' of XSec (Who? What? Where? When? Why? and How?) and argue that XSec has unique and complex characteristics: XSec
involves several different stakeholders (i.e., the system's developers, analysts, users and attackers) and is multi-faceted by nature (as it requires reasoning about system model, threat model and properties of security, privacy and trust as well as about concrete attacks, vulnerabilities and countermeasures). We define a roadmap for XSec that identifies several possible research directions.
\end{abstract}


\section{Introduction}
The security of information, data, processes, software, protocols, computers, networks and systems is notoriously a challenging problem (and very often an undecidable one). Security is difficult. It is difficult to achieve, to reason about, to apply, to understand, to teach. It is difficult to \emph{explain}.

The Defense Advanced Research Projects Agency (DARPA) recently launched the 
\begin{center}
\emph{Explainable Artificial Intelligence (XAI)} 
\end{center}
program that aims to create a suite of new AI techniques that enable end users to understand, appropriately trust, and effectively manage the emerging generation of AI systems.
Some research on explainable AI had already been published before 
DARPA's program (e.g., \cite{Swartout1991,Ye1995,Aldeco-PerezMoreau08,Aldeco-PerezMoreau10,Bidot2010,Sohrabi2011,Seegebarth2012,Papadimitriou2012,Gedikli2014}), but XAI encouraged a large number of researchers to take up this challenge. In the last couple of years, several publications have appeared that investigate how to explain the different areas of AI, such as 
\emph{machine learning}~\cite{darrell16}, \emph{recommender systems}~\cite{Muhammad2016}, \emph{robotics} and \emph{autonomous systems}~\cite{Rosenthal2016,Sheh2017,shah17}, \emph{constraint reasoning}~\cite{Freuder2017} and \emph{planning}~\cite{Chakraborti2017,FoxLongMagazzeni2017,Langley2017}.\footnote{See also \cite{Miller2017} and the other papers listed at \url{http://home.earthlink.net/~dwaha/research/meetings/faim18-xai/}.}

In this paper, inspired by the XAI program, we propose a new paradigm in security research: 
\begin{center}
\emph{Explainable Security (XSec)}. 
\end{center}
Some pioneering works on explaining security have focused on \emph{security for relational databases}~\cite{Bender2014} and on \emph{explanation and trust}~\cite{Pieters11}.
In~\cite{Bender2014}, Bender, Kot and Gehrke 
propose a new model in which policy decisions are explainable. In this model, instead of simply rejecting an unauthorized query by a principal, the system provides the principal with a concise explanation of why the query was rejected and what additional permissions the principal would need to be granted for a successful execution. The principal can then refine the query or request additional permissions based on the explanation provided. 

In~\cite{Pieters11}, Pieters investigates the relation between explanation and trust, focusing in particular on expert systems and e-voting systems. Pieters observes that
\begin{quote}\em
In artificial intelligence, explanations are usually provided by the system itself. In information security, explanations are provided by the designers. Nonetheless, in \emph{both} artificial intelligence and information security, the role of explanations consists for a major part of acquiring and maintaining the trust of the user of the system. 
\end{quote}
He discusses how explanations are required for trust:
\begin{quote}\em
Here, the question is how it is possible to communicate the analysis that experts have made of a security-sensitive system to the public. Why is it secure? Or, more appropriately: How is it secure?
\end{quote}
Explanations are thus 
\begin{quote}\em
thought to bridge the gap between `actual security’ and `perceived security'.
\end{quote}
Pieters also discusses two main goals that an explanation may have: \emph{transparency} (e.g., to allow users to understand what the designers have done to protect them) and \emph{justification} (e.g., offering reasons for an action). He contrasts \emph{explanation-for-trust} (i.e., explanation of how a system works, by revealing details of its \emph{internal} operations) with \emph{explanation-for-confidence} (i.e., explanation to make the user feel comfortable in using the system, by providing information on its \emph{external} communications).

\begin{figure*}
\begin{center}
\includegraphics[scale=0.6475]{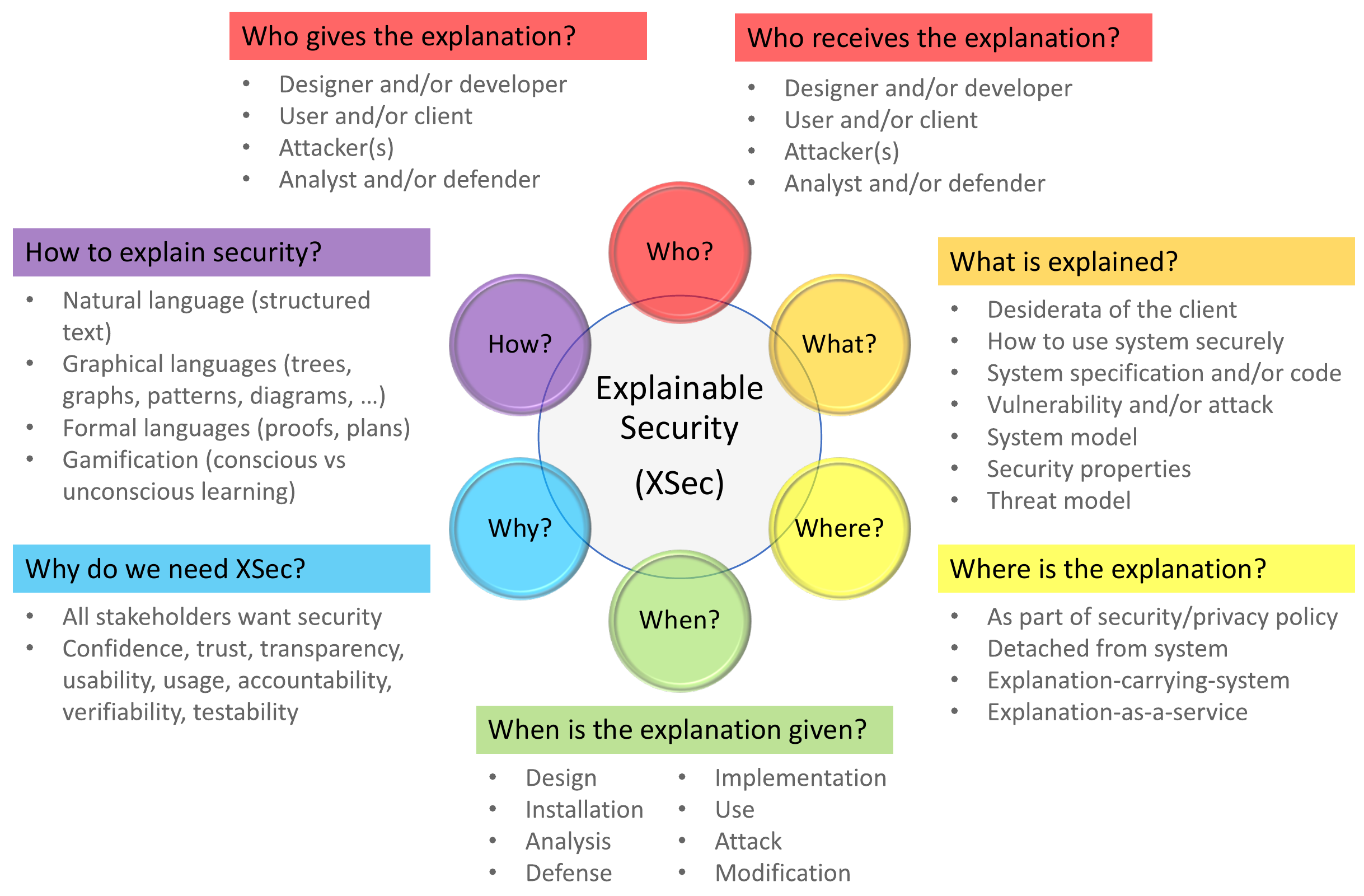}
\caption{The Six Ws of Explainable Security}
\label{fig:XSec}
\end{center}
\end{figure*}

We argue that XSec is a difficult problem and that
it has unique and complex characteristics.
In fact, XSec is more complex than what is discussed by Bender et al.~and by Pieters; it is also more complex than \emph{usable security}~\cite{usable,WashZurko2017}, \emph{security awareness}~\cite{banfield2016study,yildirim2016importance} and \emph{security economics}~\cite{Anderson}. This is because XSec involves several different stakeholders (i.e., the system's developers, analysts, users and attackers) and is multi-faceted by nature (as it requires reasoning about system model, threat model and properties of security, privacy and trust as well as about concrete attacks, vulnerabilities and countermeasures). XSec is thus an exciting novel paradigm that requires a full-fledged and heterogeneous research program to be realized. In the following, we define a roadmap that identifies several possible research directions. To describe the challenges of XSec and how they could be tackled, we proceed by discussing the ``Six Ws'' of XSec summarized in Figure~\ref{fig:XSec}: Who? What? Where? When? Why? and How?

\section{Who?}
Consider a generic \emph{system}, where we use here ``system'' to refer to a generic process, software, protocol, computer, network, cyber-physical system, critical infrastructure, etc. that processes information/data whose security must be protected, where ``security'' similarly generically refers to one or more of the security properties of interest, including confidentiality, integrity, availability, authentication, authorization, non-repudiation, accountability, unobservability, privacy, etc. 

The \emph{dramatis personae} of XSec 
are:
\begin{itemize}
\item the \emph{designer} (and/or \emph{developer}) of the system, who has designed (and/or developed) the system to guarantee a number of specified security properties;
\item the \emph{user} (and/or \emph{client}) of the system, who can typically be assumed to be an honest non-expert who might commit mistakes that make the system vulnerable;
\item the \emph{attacker} (or more \emph{attackers}) of the system, who searches for and exploits vulnerabilities of the system 
for reasons or profit, fame, reward, etc.;
\item the \emph{analyst} of the system, who carries out a semi-formal or formal analysis of the system at design time (or based on the specification of a deployed system) or tests the system at runtime (e.g., using penetration testing, vulnerability-based testing or model-based testing); 
\item the \emph{defender} of the system, who attempts to protect the system, e.g., by monitoring the activities of the system and reacting to the attacker's actions.
\end{itemize}

For some situations, the recipient of the explanation will be an agent rather than a human, and we can then contrast \emph{internal explanations} (designed for software agents) and \emph{external explanations} (designed for humans, which is what XAI research typically focuses on).

Some of 
the above roles might actually be played by the same ``principal'' (agent or human), as the designer might for instance act also as analyst or defender, the analyst might also provide immediate defense and the attacker might be a user of the system. The literature is full of examples of vulnerabilities caused by mistakes by designers or users, along with details of the corresponding attacks. Some of these attacks could have been prevented by better explanations. In fact, 
\begin{quote}
all of these roles might require explanations or need to act as explainer,
\end{quote}
For instance, 
\begin{itemize}
\item a designer and/or an analyst might need to explain to the user how to interact with the system, why the system is secure and why it carries out a particular action (in line with what is discussed in~\cite{Bender2014,Pieters11});
\item a user or client might need to explain to the designer or the analyst how he expects the system to behave and how they typically interact with the system, to allow the designer to elicit the requirements for building the system in the first place and to allow the analyst to validate the security of user interactions; 
\item an attacker might need to explain the attack strategy to his accomplices 
so that they can attack in coordination, or he might have used a complex penetration testing tool to test the system for vulnerabilities and now needs the tool to explain to him the attack trace (or attack plan or strategy) that has been identified so that he can carry out the attack for real;
\item an analyst might need to explain to the designer how to improve the system's security or to the defender how and what to defend;
\item a defender might similarly need to understand possible attack traces in order to take action against them as well as explain to the users how they should behave to protect the system and themselves.
\end{itemize}
In addition to this, it is also necessary to tackle the research question of what actually constitutes a good (and secure) explanation, as we discuss in more detail in the following sections.

Before we do so, let us consider a concrete example that arises from the observation that when dealing with sensitive data, classical authentication solutions based on username-password pairs are not enough. The ``General Data Protection Regulation''~\cite{directive} mandates that specific security measures must be implemented, including \emph{multi-factor authentication (MFA)}, an authentication solution that aims to augment the security of the basic username-password authentication by exploiting two or more authentication factors (see, e.g., \cite{SCRV2018}). 
In~\cite{ecb_strong}, MFA is for instance defined as:
\begin{quote}\em\small
``a procedure based on the use of two or more of the following elements --- categorised as knowledge, ownership and inherence: i) something only the user knows, e.g., static password, code, personal identification number; ii) something only the user possesses, e.g., token, smart card, mobile phone; iii) something the user is, e.g. a biometric characteristic, such as a fingerprint. In addition, the elements selected must be mutually independent [...] at least one of the elements should be non-reusable and non-replicable''.
\end{quote}
The underlying idea is that the more factors are used during the authentication process, the more confidence a service has that the user is correctly identified. This is the basic explanation provided by the designer to the user to justify a more involved authentication that the user might perceive as cumbersome. However, the user might also need to be told that choosing a weak password is a bad idea even in the case of MFA. Two attacker accomplices who carry out a coordinated attack against the two components of MFA might need to explain their sub-attack to each other to ensure their ultimate success. The analyst who has discovered an attack to the MFA system might need to explain to the designer why the attack succeeeded and how to patch it. The analyst/defender might also need to explain to the users why they should, e.g., abandon the use of one of the elements they had been using so far and switch to using another pair of elements; for instance, because the new password that a user has chosen is too weak and thus easily guessable, or because the device on which the user is trying to authenticate does not include a biometric reader.

There are thus many things to explain by/to many different stakeholders, which is one of the main reasons why XSec, even in the case of a relatively simple example such as MFA, is a challenging endeavor. In the following, we discuss the five remaining Ws, although we have already anticipated much of the discussion here (which is somewhat unavoidable given that the Ws are not independent but quite deeply intertwined).


\section{What?}
It is not enough to explain the system in a generic way. First of all, the different stakeholders will need explanations at different levels of detail and with different aims: 
\begin{itemize}
\item designers/developers will need an explanation of the desiderata of the client that is detailed enough for them to be able to realize the system in a satisfactory and secure way (e.g., if the client wants a system that replaces passwords with face recognition);
\item non-expert users will need an explanation that increases their confidence and trust, and that also teaches them how to use the system correctly and securely (e.g., if passcodes are used as back-up in case face recognition fails as in the iPhone X, then the user should be made aware that the passcode ought to be strong enough and not guessable such as a date of birth or a phone number, since otherwise an attacker who steals the iPhone X will obviously fail face recognition but the iPhone X will allow him to get access by guessing the passcode);
\item analysts will need access to the system's specification or to the system's code in order to be able to create a model to analyze or to be able to generate and execute test cases;
\item designers/developers/defenders will need an explanation of a vulnerability and related attacks in order to implement patches or  defenses;
\item attackers will need an explanation of how to exploit possible vulnerabilities, of why their attacks failed and of the implications of new security techniques on their attack strategies.
\end{itemize}

Second, several different ``things'' will need to be explained, including:
\begin{itemize}
\item the \emph{system} and the \emph{system model} used for design, implementation and analysis, e.g., the model of how MFA actually works;
\item the \emph{security properties} that the system should guarantee. e.g., the authentication provided by MFA can be used as a basis to provide authorization, integrity, confidentiality, non-repudiation and so on;
\item the \emph{threat model} that has been considered by designers, developers and analysts, highlighting, in particular, the fact that a system might be secure against one threat model but insecure against another (e.g., a system might be secure against an outside attacker but insecure against insider attacks) or the fact that the successful MFA of a user won't prevent the system from being attacked when that user turns out to be malignant and reveals, say, trade secrets of the company he works for;
\item the actual \emph{vulnerability} and related \emph{attack} that has been discovered will need to be explained to the attacker (especially when the attacker used a tool to search for an attack and now needs to carry out it concretely), along with the costs and benefits of the attack;
\item the possible \emph{countermeasures} for the discovered \emph{vulnerability} and related \emph{attack} will need to be explained (along with the vulnerability and the attack) to the honest stakeholders who will need to understand the attack's impact, its risk and the mitigation strategies.
\end{itemize}
To that end, it will be helpful to answer a number of questions, including:
\begin{itemize}
\item What is actually secure? Which parts of the system? Which security properties are guaranteed and for how long? (For instance, authentication is typically granted for a session, which expires after some amount of time.) Which features are insecure and why and how can they be attacked? Are there different levels of security (e.g., for users with different rights)?
\item What is the threat model considered? Does it include insiders and outsiders? Who are the potential attackers and what do they want? Why do they want it?
\item How does the attack look like and how ``difficult'' is it? How expensive is it? (It does not make sense to use a one million dollar machine to mount an attack with a loot of a few thousand dollars.) How long will the attack take? (If students try to steal the questions of their next exam but their attack takes so long that they get hold of the questions only after the exam has been given by the professors, then there is actually no point.) This requires reasoning quantitatively about the economics of the attack (including costs, performance, time) but also about the trade-off between attacking and the risk of being discovered.
\item Under which assumptions and conditions is the system assumed to be operating securely or has been proved to be secure? For instance, many security analyses (e.g., of protocols or web applications) typically assume a \emph{Dolev-Yao-style attacker}~\cite{DolevYao1983} who controls the network but cannot break cryptography, which is quite a strong assumption to make as cryptography might indeed be broken (by classical computers and even more so by  quantum computers if and when they will be realized in their full capacity); on the other hand, relaxing this assumption and considering an attacker who might be able to break cryptography typically complicates the analysis (the problem is undecidable anyway) and the ability of an analyst to prove security guarantees.
\item What are the \emph{legal implications} of the explanation? Is the explanation ``binding''? This would require, for instance, explaining how the system works and what is expected of the user, possibly including a digital signature to acknowledge the receipt and understanding of the explanation. In case of an attack, this would also require explaining what happened and why, and what countermeasures can be taken (and by whom).
\end{itemize}

\section{Where?}

We have already addressed the question of which ``parts'' of the system need to be explained in the ``What?'' section. Now we focus briefly on the question of where the explanations should be made available. A number of different options are available here, including the following four main ones:
\begin{itemize}
\item One could include the explanations to the users as part of the security/privacy policy, but it is well known that users typically ignore the policy and scroll down as quickly as possible so that they can get on with their interaction with the system.
\item One could completely detach the explanation from the system, e.g., by making it available on a different webpage, but it is unclear to us if and how the relevant stakeholders will be made aware of where to find the explanation and whether they will decide to trust it.
\item One could consider a sort of \emph{explainable security as a service}, where stakeholders interact with an expert system to obtain and/or provide explanations.
\item One could proceed in the style of \emph{proof-carrying code}~\cite{Necula1997}, ``appending'' a possibly digitally-signed explanation to the system to achieve a \emph{security-explanation-carrying-system}. We believe that this is the most promising direction, but it will of course require considerable work to protect the explanation from attacks and actually explain to the stakeholder how they can access it and make use of it.
\end{itemize}

\section{When?}

We want Explainable Security and we want it now! Jokes aside, the many vulnerabilities that are reported daily, including some of our most widespread and supposedly secure systems (consider, e.g., recent attacks against: TLS; PGP; processors; dropbox, one drive, iCloud and other cloud systems; biometric authentication systems; e-commerce and e-banking systems; e-voting systems, etc.), are witness to the fact that security is indeed difficult to achieve (which is why security has been and still is one of the hottest research topics) but also that in many cases security systems are difficult to explain to the different stakeholders.

We need to explain security when the system is 
\begin{itemize}
\item designed,
\item implemented,
\item deployed and installed,
\item used, 
\item analyzed,
\item attacked,
\item defended,
\item modified,
\end{itemize}
and possibly even when the system is decommissioned and replaced, so that the different stakeholders understand why this decision was taken and how the new system will improve over the old one. 

In particular, explanations will need to be defined and provided at \emph{design time} (when the system is developed) but also at \emph{runtime} (when the system is running). For the runtime case, think, e.g., of a critical system, critical infrastructure or cyber-physical system such as a nuclear power plant in which a supervisor is in charge of setting high/low security levels and of intervening in the case of an ongoing attack to estimate the success chances of the attack, understand its impact on the system and adopt possible countermeasures (see, e.g., \cite{LanotteMMV17}). The attack could have disastrous consequences (e.g., manipulating the SCADA and PLC systems of a power plant as the Stuxnet Worm~\cite{stuxnet}) or (appear to) be non-threatening as it manipulates sensors and actuators of the system but without bringing them outside of their tolerance zone so that the supervisor actually decides not to intervene.

\section{Why?}

This is easiest question to answer: because all the different stakeholders of a system want it to be secure (well, with the exception of the attacker, of course). Explanations will help increase confidence, trust, transparency, usability and concrete usage (in the sense that users will be more keen to adopt the system), accountability, verifiability and testability.

\section{How?}

As we already remarked above, the different stakeholders will need explanations at different levels of detail and with different aims, and these explanations will need to be comprehensible, timely and accurate (among other properties).
The explanations will need to be written in a language (and with a description strategy) suitable for the intended audience, including 
\begin{itemize}
\item \emph{natural language} (used to produce informal but possibly structured text written in English or any other language understandable by the audience);
\item \emph{graphical languages} such as explanation trees, attack trees, attack-defense trees, attack graphs, attack patterns, message-sequence charts, use case diagrams ...;
\item \emph{formal languages} including proofs and plans;
\item games that have been produced as the result of a \emph{gamification} process to teach users how to interact with a system (although one could actually object that such games often provide for some ``unconscious learning'' in which the user learns how to interact but without really understanding why).
\end{itemize}
It should also be investigated whether one learns more by seeing a proof of the security of the system or by being shown an attack against an insecure system. Both are of course useful, but they explain different things in a different way, and they can both be traced back to the question asked by Pieters~\shortcite{Pieters11} of ``how it is possible to communicate the analysis that experts have made of a security-sensitive system to the public''.

It will also be necessary to evaluate security explanations originating from applications of XAI and other areas of computer science, and possibly even social sciences, psychology and other disciplines. Careful subject studies will need to be designed to assess measurement categories such as: a priori measures of explanation quality, user satisfaction, mental model understanding, and user-machine task performance. 

Moreover, the explanation processes will themselves have to be designed properly, tested thoroughly and deployed correctly, and it will be useful to investigate the trade-off between such explanation processes and the security threats.



We expect that many of the How? questions posed in the specific case of security will actually be answerable by suitably adapting and extending the techniques and tools that have been and are being developed for XAI.
Still, we conclude the discussion of the Ws by considering again the claim that we made above that XSec has unique and complex characteristics, 
and is more challenging than the pioneering research on explanations in security~\cite{Bender2014,Pieters11}. Let us illustrate this by an example that shows that XSec calls for proper extensions of the research on XAI and for novel investigations.

In Explainable Planning~\cite{FoxLongMagazzeni2017}, one of the questions the planner should answer is why things cannot be done and why and how one needs to replan. Similarly, in the model of Bender et al., the relational database system provides the principal with a concise explanation of why the query was rejected and what additional permissions the principal would need to be granted for a successful execution. If one considers a more general security system, however, such an explanation might make the system less secure! This is because the explanation itself might reveal security-sensitive information.

For instance, the attacker might not know whether a certain person is indeed a user of the system: trying to login pretending to be that user and being told that the user does not exist, or that the password is wrong, or that the user needs more privileges to be able to carry out some specific action already constitutes a leak of information.\footnote{As another example, consider argument passing within GET protocols in HTTP requests, which basically allow the attacker to understand which is the name of the server's variable that handles the request, thus facilitating the attack. POST protocols are to be preferred, but they can't be used always.}
Hence, explanations need to be ``relativized'' and in some cases made less ``powerful'' by withholding certain details. But a less powerful explanation is essentially an incomplete explanation, which will be ignored or not fully achieve its purpose. The quest for a reasonable trade-off thus makes XSec particularly challenging.











\section{Conclusions}

We see XSec as a new paradigm that brings together many hot topics and current trends but also indicates the need of exploring unchartered territory.
This paper has just skimmed the surface of XSec by pointing out some of the main objectives and challenges, and defining a roadmap that identifies several 
possible research directions. 

We have already begun a more detailed investigation of the different techniques and tools that can be exploited or need to be developed to answer the questions posed by the different Ws, as well as formalizing the relationships and interdependences between the Ws. More specifically, we expect that, using the growing amount of results on XAI together with the pioneering research on explanations in security and trust as a stepping stone, we will soon be able to provide some concrete answers. To that end, we will also take inspiration from research that have attempted to ``bridge the gap'' between different research questions and different communities, such as the work of Hoffman on planning for penetration testing~\shortcite{Hoffmann2015}. 

This is the beginning of a beautiful friendship between explanations and security, and we plan to be part of the fellowship that nurtures this friendship.



\section*{Acknowledgments}
We thank David W.~Aha and the anonymous reviewers of the ``IJCAI/ECAI 2018 Workshop on Explainable Artificial Intelligence (XAI)'' for their very useful comments and suggestions.


\end{document}